\documentclass[12pt]{scrartcl}
\usepackage{amssymb,amsmath,bm,bbm}
\usepackage{epsf}
\usepackage{epsfig}
\usepackage{afterpage}
\usepackage{longtable}
\usepackage{cite}
\usepackage{latexsym, mathrsfs}
\usepackage{graphics}
\usepackage{url}
\usepackage{paralist}
\usepackage{bbold}
\usepackage{fancyhdr}
\advance \headheight by 3.0truept       

\begin{document}

\begin{titlepage}
\vspace*{-0.5truecm}

\begin{flushright}
{FLAVOUR(267104)-ERC-27}
\end{flushright}

\begin{center}
\boldmath
{\Large\textbf{ Correlations in Minimal $U(2)^3$ models and an  SO(10) SUSY GUT model facing new data}}
\unboldmath
\end{center}

\begin{center}
{\bf   Jennifer Girrbach}
\vspace{0.3truecm}

{\footnotesize
{\sl TUM Institute for Advanced Study, Technische Universit\"at M\"unchen, \\ D-85748 Garching, Germany
}

}

\end{center}

\begin{abstract}
{\small\noindent

Models with an approximate $U(2)^3$ flavour symmetry represent simple non-MFV extensions
of the SM. We compare correlations of $\Delta F = 2$ observables in CMFV and in a minimal version of $U(2)^3$ models, $MU(2)^3$,
where only the minimal set of spurions  for breaking the symmetry is used and where only SM operators are relevant. Due to the
different treatment of the third generation $MU(2)^3$ models avoid the $\Delta M_{s,d}-|\varepsilon_K|$ correlation of CMFV which
precludes to
solve the $S_{\psi K_S}- |\varepsilon_K|$ tension present in the flavour data. While the
flavour structure in $K$ meson system is the same for CMFV and $MU(2)^3$ models, CP violation in $B_{d,s}$ system can deviate
in $MU(2)^3$ models from CMFV. We point out a triple correlation between $S_{\psi\phi}$, $S_{\psi K_S}$ and $|V_{ub}|$ that can
provide a distinction between different $MU(2)^3$ models.

GUTs open the possibility to transfer the neutrino mixing matrix $U_\text{PMNS}$ to the quark sector which leads to correlations
between leptonic and hadronic observables. This is accomplished in a
controlled way in an SO(10) SUSY GUT model proposed by Chang, Masiero and Murayama (CMM model) whose flavour structure differ
significantly from the constrained MSSM.  We present a summary of a global analysis of several flavour processes containing
$B_s-\overline{B}_s$ mixing, $b\to s\gamma$ and $\tau\to\mu\gamma$. Furthermore we comment on the implications on the model due
to the latest data of $S_{\psi\phi}$, $\theta_{13}$ and the Higgs mass.}
\end{abstract}
\end{titlepage}

\section{Current situation of the flavour data}

\begin{figure}[!tb]
   \centering 
\includegraphics[width = 0.5\textwidth]{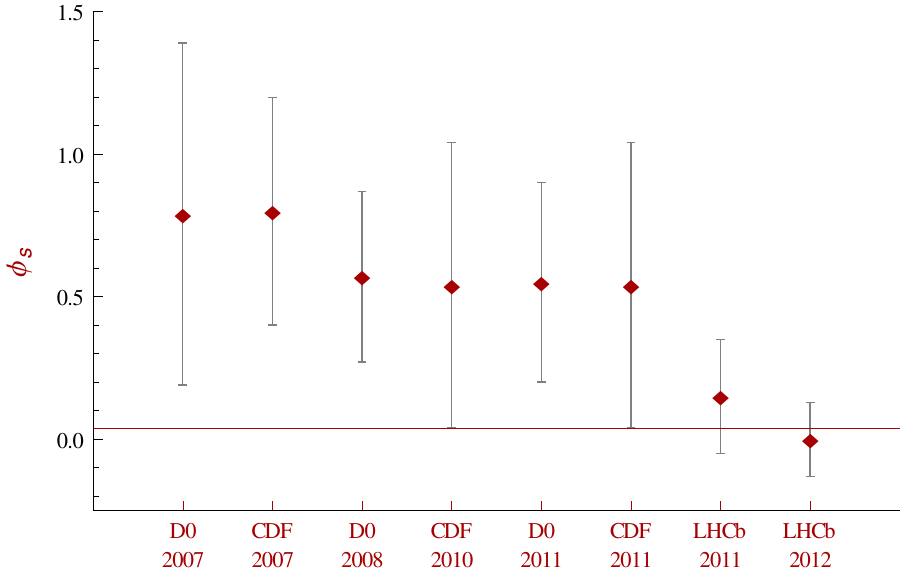}
\includegraphics[width = 0.45\textwidth]{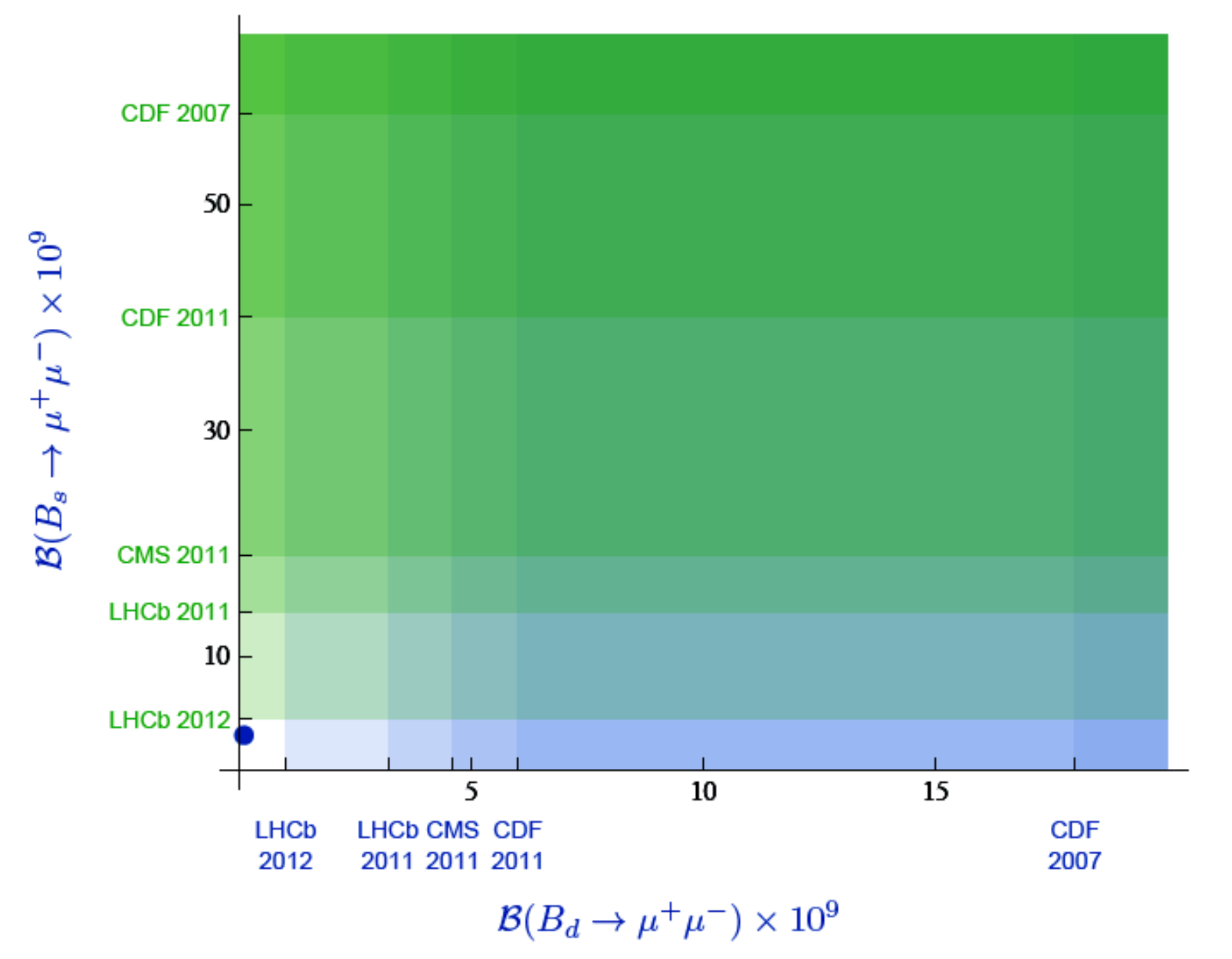}
\caption{Left: Measurements of the CP phase $\phi_s$ from D\O, CDF and LHCb. Right: Upper bounds on $\mathcal{B}(B_{s,d}\to
\mu^+\mu^-)$ where the blue dot corresponds to the central value of the SM prediction
\cite{Aaij:2012ac,Aaij:2012qe,LHCb-CONF-2012-002}.}\label{fig:LHCbdata}~\\\hrule
\end{figure}

With the start of the LHCb experiment a new era in precision measurements in flavour physics started.
The present 95\% C.L. upper bound  $\mathcal{B}(B_s\to \mu^+\mu^-)\leq 4.5\cdot 10^{-9}$ \cite{Aaij:2012ac} is already close to
the
Standard model (SM) prediction  $\mathcal{B}(B_s\to \mu^+\mu^-)^\text{SM} = (3.1\pm 0.2)\cdot 10^{-9}$
\cite{Buras:2012ts,LHCb:2012py}\footnote{In \cite{Buras:2012ru} the ``non-radiative'' branching ratio that corresponds to the
branching ration fully inclusive of bremsstrahlung radiation was calculated to  $\mathcal{B}(B_s\to \mu^+\mu^-) =(3.23 \pm 0.27)
\cdot 10^{-9}$.}. When the corrections from $\Delta \Gamma_s$, pointed out in \cite{deBruyn:2012wj,deBruyn:2012wk} are taken into
account the experimental upper bound is reduced to $4.1\cdot 10^{-9}$. New data on mixing induced CP
violation in $B_s-\overline{B}_s$ mixing measured by $S_{\psi\phi} = 0.002\pm 0.0087$ \cite{Clarke:1429149} is consistent with the SM
prediction of $S_{\psi\phi}^\text{SM} =0.0035\pm 0.002 $  and excludes ranges from CDF and D\O\ with large $S_{\psi\phi}$.
Thus there is not much room left for new physics (NP).
The  experimental situation is displayed in Fig.~\ref{fig:LHCbdata}\footnote{I thank Maria Valentina Carlucci for providing me
these two plots.}.

\begin{figure}[!tb]
  \centering
\includegraphics[width = 0.5\textwidth]{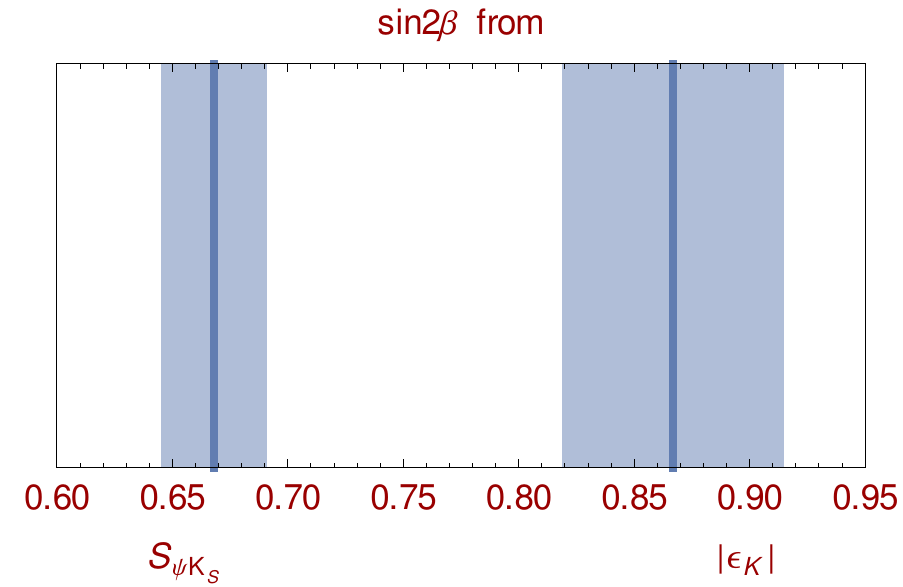}
\caption{ $\sin2\beta$ determined from $S_{\psi K_S}$ (left) and $|\varepsilon_K|$ (right).}\label{fig:sin2beta}~\\\hrule
 \end{figure}

However a slight tension in the flavour data concerns $|\varepsilon_K|$, $B^+\to \tau^+\nu$ and $S_{\psi K_S}$ which can be
related with the so-called
$|V_{ub}|$-problem. In the SM $S_{\psi K_S}$ measures the angle $\beta$ of the unitarity triangle directly: $S_{\psi K_S}  = \sin
2\beta$. Due to
$|\varepsilon_K|\propto \sin2\beta |V_{cb}|^4$ both quantities are correlated in the SM (but the $|V_{cb}|^4$ dependence leads to
additional
uncertainties). This issue was discussed in \cite{Lunghi:2008aa,Buras:2008nn} and a $ 3.2\sigma$ discrepancy was identified in
2008. However this tension went down to about $2\sigma$. In
Fig.~\ref{fig:sin2beta} one can see that the $\sin2\beta$ derived from the experimental value $S_{\psi K_S}$ is
much smaller that the one derived from $|\varepsilon_K|$. The ``true'' value of $\beta$~-- the angle opposite of the $|V_{ub}|$-side of the
unitarity triangle~-- depends on the value of $|V_{ub}|$ and $\gamma$.  However there is a tension between the exclusive and
inclusive determinations of  $|V_{ub}|$ \cite{Nakamura:2010zzi}:
\begin{align}
 &|V_{ub}^\text{incl.}| = (4.27\pm 0.38)\cdot 10^{-3}\,,\qquad |V_{ub}^\text{excl.}| = (3.38\pm 0.36)\cdot
10^{-3}\,.
\end{align}
 Now one can distinguish between these two benchmark scenarios: If one uses the
exclusive (small) value of $|V_{ub}| $ to derive $\beta_\text{true}$ and then calculates $S_{\psi
K_S}^\text{SM}=\sin2\beta_\text{true}$ one
finds agreement with the data whereas $|\varepsilon_K|$ stays below the data. Using the inclusive (large) $|V_{ub}| $ as input for 
$\beta_\text{true}$  the predicted $S_{\psi K_S}$ is above the measurements while $|\varepsilon_K|$ is in agreement with the
data. However in such considerations one has to keep in mind the error on  $|\varepsilon_K|$ coming dominantly from the 
error of $|V_{cb}|$ and the error of the QCD factor $\eta_1$ in the charm contribution~\cite{Brod:2011ty}.

The branching ratio $\mathcal{B}(B^+\to \tau^+\nu)\propto F_{B^+}^2|V_{ub}|^2$ can also be used to  measure $|V_{ub}|$. The SM
prediction  $\mathcal{B}(B^+ \to \tau^+ \nu)_{\rm SM}= (0.80 \pm 0.12)\cdot
10^{-4}$ as calculated in \cite{Altmannshofer:2009ne} where  one eliminates
the uncertainties of $F_{B^+}$ and $|V_{ub}|$ by using $\Delta M_d$,  $\Delta M_d/\Delta M_s$ and
$S_{\psi K_S}$  is about a factor 2 below the experimental world average based
on results by BaBar \cite{Aubert:2009wt} and Belle \cite{Ikado:2006un}: $\mathcal{B}(B^+ \to \tau^+ \nu)_{\rm exp} = (1.67 \pm
0.30) \cdot 10^{-4}$ \cite{HFAG}. Consequently this favors a large $|V_{ub}|$ and leads to a $S_{\psi
K_S}-\mathcal{B}(B^+\to\tau^+\nu)$ tension discussed for example in \cite{Nierste:2011na}. Recently  new results have been
provided by
BaBar $\mathcal{B}(B^+ \to \tau^+ \nu)_{\rm exp} = (1.79 \pm 0.48) \cdot 10^{-4} $ \cite{Lees:2012ju}
and by Belle $
\mathcal{B}(B^+ \to \tau^+ \nu)_{\rm exp} = (0.72 \pm^{0.27}_{0.25} \pm^{0.46}_{0.51}) \cdot 10^{-4}$
\cite{Adachi:2012mm} where the latter   value went down and is consistent with the SM prediction.

It is now interesting to see if a certain new physics model can solve these problems and if yes, which $|V_{ub}|$ scenario is
chosen.
In the following we will confront constraint minimal flavour violation (CMFV) and models with a global $U(2)^3$ symmetry to this
tension. At the end we discuss a concrete SO(10) SUSY GUT model which has a different flavour structure and  can
be seen as an alternative to MFV.

\boldmath
\section{Correlations of $\Delta F = 2$ observables: CMFV vs. $MU(2)^3$ and the role of $|V_{ub}|$}
\unboldmath

\begin{figure}
\centering
 \includegraphics[width = 0.55\textwidth]{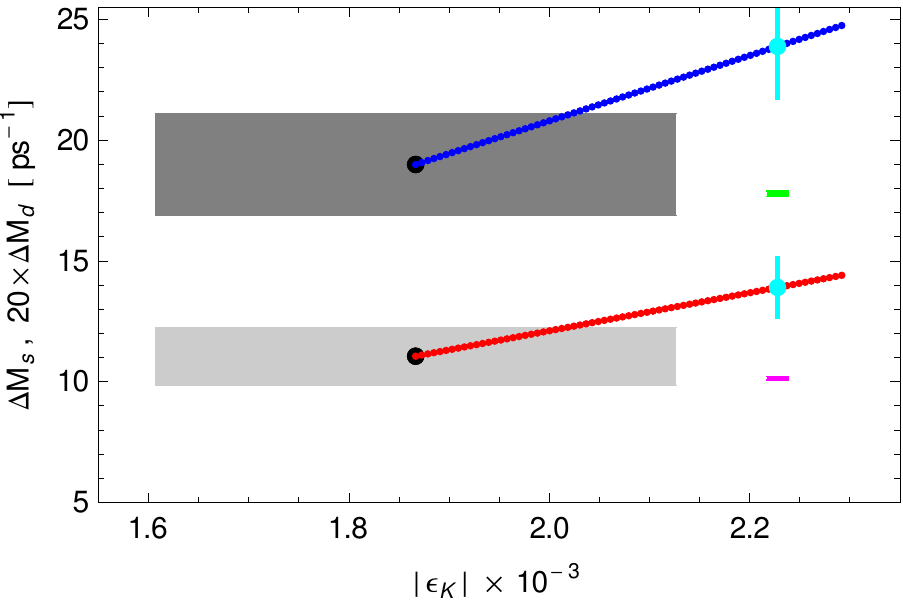}
\caption{$\Delta M_{s}$ (blue) and $20\cdot\Delta M_{d}$ (red) as functions of $|\varepsilon_K|$ in models 
with CMFV for $|V_{ub}| = 0.0034$ chosen by these models. The short green and magenta  lines represent the data, while the large
gray regions corresponds to the SM 
predictions \cite{Buras:2012ts}.}\label{fig:DeltaMvsepsKv2}~\\\hrule
\end{figure}

The great success of the Cabibbo Kobayashi Maskawa mechanism puts strong constraints on the flavour structure of NP models.
A very simple extension of the SM is CMFV, where the CKM matrix is the only source of flavour and CP violation and only SM
operators are
relevant below the electroweak scale.
Phenomenological consequences of CMFV concerning $\Delta F = 2$ observables are the following:
\begin{itemize}
 \item Since there are no new CP violating phases the mixing induced CP asymmetries stay as in the SM:
\begin{align}
 S_{\psi K_S} = \sin2\beta\,,\qquad S_{\psi\phi} =
\sin2|\beta_s|\,.
\end{align}
\item $\Delta M_{s,d}$ and $|\varepsilon_K|$ can only be enhanced relative to the SM and this enhancement is
correlated \cite{Blanke:2006yh,Buras:2000xq}.
\item   CMFV chooses exclusive $|V_{ub}|$ because $S_{\psi K_S}$ stays as in the SM and $|\varepsilon_K|$ can be enhanced.
But if one wants to solve the $|\varepsilon_K|-S_{\psi K_S}$ tension one gets a problem with $\Delta M_{s,d}$. This
$\Delta M_{s,d}-|\varepsilon_K|$ tension is shown in
Fig.~\ref{fig:DeltaMvsepsKv2}.
\end{itemize}

\begin{figure}[!tb]
 \centering
 \includegraphics[width = 0.45\textwidth]{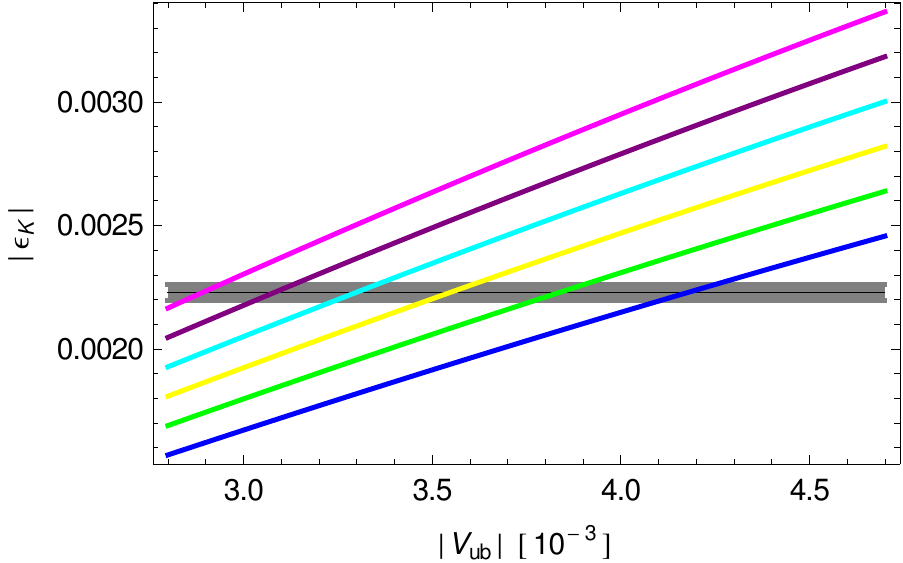}
 \includegraphics[width = 0.45\textwidth]{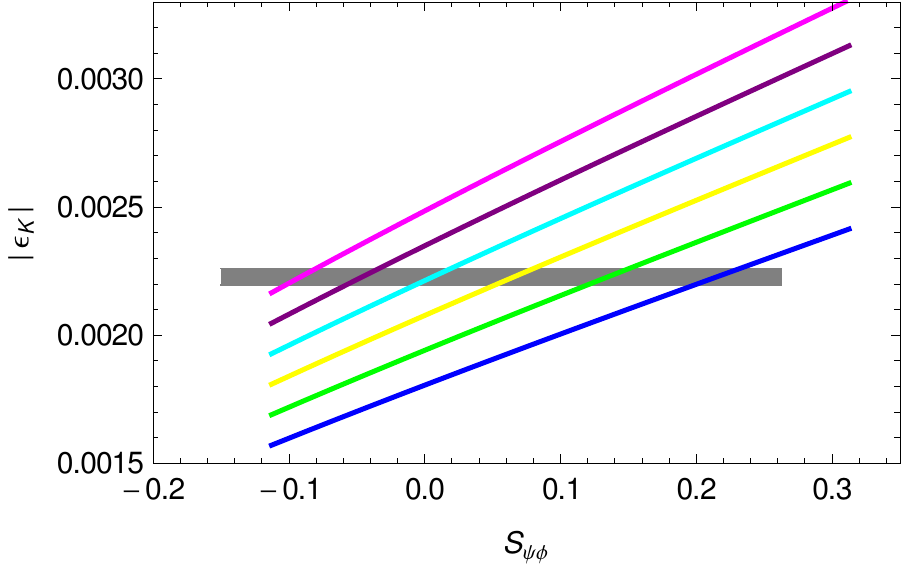}
 \caption{$|\varepsilon_K|$ as a function of $|V_{ub}| $ in  models with 
 $U(2)^3$ symmetry  (left) and $|\varepsilon_K|$
versus $S_{\psi \phi}$ (right) for fixed $S_{\psi K_S}=0.679$ and $|V_{ub}|\in[0.0028,0.0046]$ and different 
 values of the enhancement factor $r_K$. From top to bottom: $r_K = 1.5$ (magenta), $1.4$ (purple), $1.3$ (cyan), $1.2$ (yellow),
 $1.1$
 (green), $1$ (blue, SM prediction)). Gray  region: experimental $3\sigma$ range of $|\varepsilon_K|$.
}\label{fig:epsK1}~\\\hrule
 \end{figure}

Consequently, the solution of the $|\varepsilon_K|-S_{\psi K_S}$ tension in CMFV shifts the problem to $\Delta M_{s,d}$.
Models with a global $U(2)^3$ flavour symmetry represent simple non-MFV extensions of the SM and can help avoiding this $\Delta
M_{s,d}-|\varepsilon_K|$ tension 
of CMFV. In these $U(2)^3$ models the stringent correlations between flavour observables in CMFV are relaxed as the third
generation is treated differently without loosing too much of its predictive capability.
The $U(2)^3$ symmetry was first studied in 
\cite{Pomarol:1995xc,Barbieri:1995uv} and then in
 \cite{Barbieri:2011ci,Barbieri:2011fc,Barbieri:2012uh,Barbieri:2012bh,Crivellin:2011fb,Crivellin:2011sj,Crivellin:2008mq}
where  a detailed description of the model can be found. 
In a minimal version of this model the global flavour symmetry $G_F = U(2)_Q\times U(2)_u\times U(2)_d$ (short: $U(2)^3$) is
broken minimally by three spurions
\begin{align}\label{equ:breakingpattern}
 \Delta Y_u = (\mathbf{2}, \mathbf{\overline{ 2}}, 1)\,,\quad\Delta Y_d =  (\mathbf{2}, 1, \mathbf{\overline{ 2}})\,,\quad V = 
(\mathbf{2}, 1, 1)\,.
\end{align}
This symmetry can  be motivated by the observed pattern of quark masses and mixings which cannot be explained in MFV models
based on a $U(3)^3$ symmetry.
A nice feature of $U(2)^3$ is that one can easily embed Supersymmetry (SUSY) with heavy 1$^\text{st}$/2$^\text{nd}$ sfermion
generation and a light
3$^\text{rd}$
generation which is still consistent with current collider bounds on sparticle masses. For more details of the model see the talk
by
Filippo Sala during this workshop \cite{Salatalk}.
General consequences of $U(2)^3$ and the breaking pattern in~(\ref{equ:breakingpattern}) concerning $\Delta F = 2$ observables
are the following:
\begin{itemize}
 \item The flavour structure in the $K$-meson system is governed by MFV (no new phase $\varphi_K$).
\item Corrections in $B_{d,s}$ system are proportional to the CKM structure of the SM and they are universal: $C_{B_d} = C_{B_s}
=: r_B$.
\item There exists one new universal phase that only appears in $B_{d,s}$ system: $\varphi_d = \varphi_s =: 
\varphi_\text{new}$.
\end{itemize}
 \begin{figure}[!tb]
  \centering
 \includegraphics[width = 0.6\textwidth]{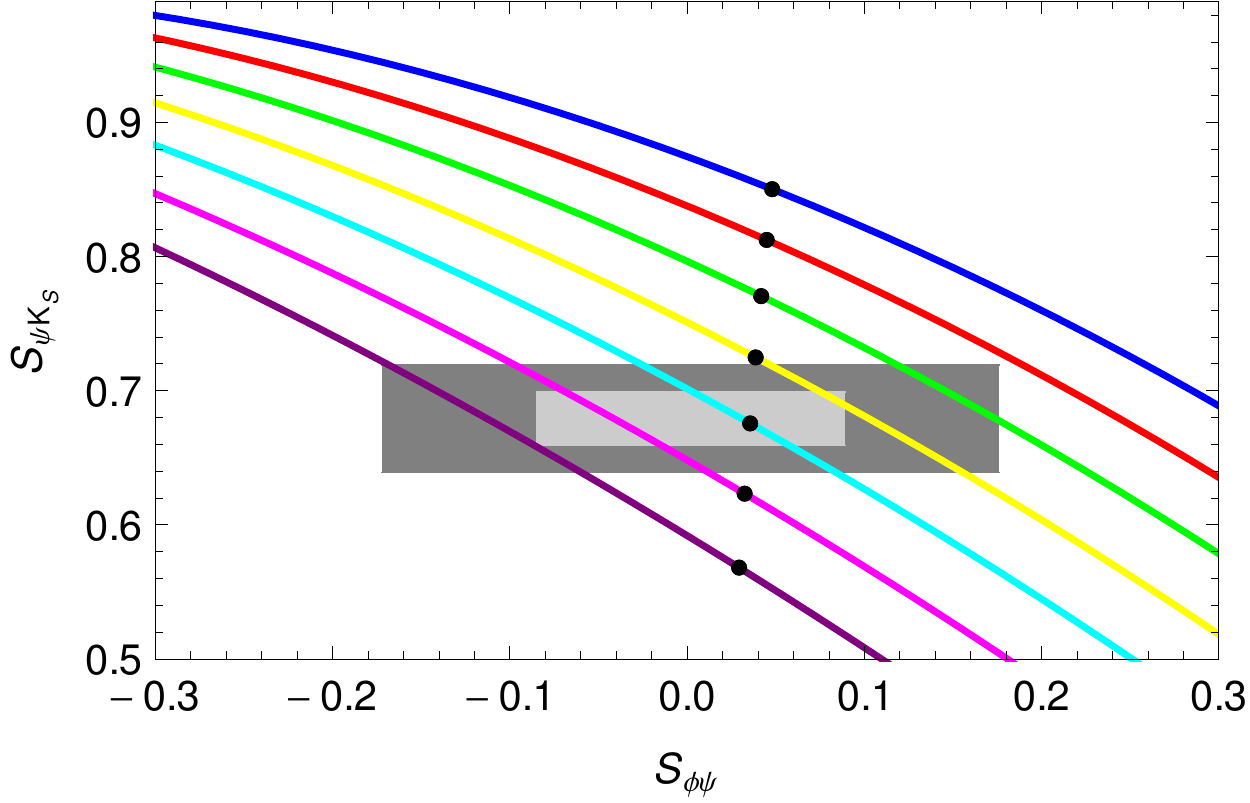}
 \caption{  $S_{\psi K_S}$ versus $S_{\psi \phi}$ in  models with 
 $U(2)^3$ symmetry for different values of $|V_{ub}|$. From top to bottom: $|V_{ub}| =$ $0.0046$ (blue), $0.0043$ (red), $0.0040$
(green),
 $0.0037$ (yellow), $0.0034$ (cyan), $0.0031$ (magenta), $0.0028$ (purple). Light/dark gray: experimental $1\sigma/2\sigma$
 region \cite{Buras:2012sd}.}\label{fig:SvsS}~\\\hrule
\end{figure}
If we further assume that only SM operators are relevant we call it minimal $U(2)^3$: $MU(2)^3$.
These properties lead to the following equations describing $\Delta F = 2$ observables where only three new parameters appear
\begin{align}
& S_{\psi K_S} = \sin(2\beta + 2 \varphi_\text{new})\,,\qquad
S_{\psi\phi} = \sin(2|\beta_s|-2\varphi_\text{new})\,,\\
&\Delta M_{s,d} = \Delta M_{s,d}^\text{SM} r_B\,,\qquad \qquad\quad \varepsilon_K =
r_K\varepsilon_K^\text{SM,tt} + \varepsilon_K^\text{SM,cc+ct}\,.
\end{align}
The parameters $r_{K,B}$ are real and positive definite and further $r_K\geq 1$. In contrast to CMFV $r_B$ and $r_K$ are in
principle unrelated. However in concrete realizations of the model, e.g. SUSY
they are correlated since they both depend on SUSY masses.
In \cite{Buras:2012sd} we point out a  triple 
$S_{\psi K_S}-S_{\psi\phi}-|V_{ub}|$ correlation which will provide a crucial test of the
$MU(2)^3$ scenario once the three observables will be precisely known. This is shown in Fig.~\ref{fig:SvsS} for fixed $\gamma
= 68^\circ$. Varying $\gamma$ between $63^\circ$ and $73^\circ$ does not change the result significantly.
Negative $S_{\psi\phi}$ is for example only possible for small $|V_{ub}|$ in the ballpark of the exclusive value. 
For inclusive $|V_{ub}|$, $S_{\psi\phi}$ is always larger than the SM prediction. $MU(2)^3$ models that are consistent with this
correlation should also describe the data for $|\varepsilon_K|$ and $\Delta M_{d,s}$. For example for $S_{\psi\phi}<0$
the particular $MU(2)^3$ model must provide a 25\% enhancement of $|\varepsilon_K|$ (see Fig.~\ref{fig:epsK1} left plot).
Moreover, if this $MU(2)^3$ flavour symmetry turns out to be true one can determine $ |V_{ub}|$ by means of precise measurements
of $S_{\psi K_S}$ and $S_{\psi \phi}$ with small hadronic uncertainties.

The dependence of $|\varepsilon_K|$ (only central values) on $|V_{ub}|$ for different values of $r_K$ is shown in the left plot of
Fig.~\ref{fig:epsK1}. Here we can read off that for a low value of $|V_{ub}|$ an enhancement of  $|\varepsilon_K|$  is needed.
Fixing $S_{\psi K_S}=0.679$ to its central experimental value we can use the triple correlation to get the connection between
$|\varepsilon_K|$ and $S_{\psi \phi}$ which is shown in the right plot of Fig.~\ref{fig:epsK1}. Thus we see that even in
$MU(2)^3$ models correlations between $B$- and $K$-physics are possible. 

\section{SO(10) SUSY GUT: CMM model }

In an SO(10) SUSY GUT model proposed by Chang, Masiero and Murayama \cite{Chang:2002mq} and also by Moroi \cite{Moroi:2000tk} the
neutrino mixing matrix $U_\text{PMNS}$ is transfered to the right-handed down quark and charged lepton sector which can induced
additional flavour violation at an observable level. In \cite{Girrbach:2011an} we have performed a global analysis in the CMM
model including an extensive renormalization group (RG) analysis to connect Planck-scale and low-energy parameters. A short
summary of this work can also be found in \cite{Buras:2012ts,Girrbach:2011wt,Nierste:2011na}. In view of the new 
knowledge about the Higgs mass and the latest measurements of the reactor neutrino mixing angle $\theta_{13}$ an updated analysis
of this model would be desirable.

\subsection{Flavour structure}

The basic ingredient of the flavour structure of the CMM model is that not only the neutrinos are rotated with $U_\text{PMNS}$
but the whole $\mathbf{5}$-plets of SU(5) $\mathbf{5}_i = (d_{Ri}^c, \,\ell_{Li},\,-\nu_{\ell_i})^T$. Whereas mixing of
right-handed quark fields
in flavour space is unphysical it is not for the corresponding superfields due to the soft breaking terms. Consequently the large
atmospheric neutrino mixing angle $\theta_{23}\approx 45^\circ$ is responsible for large $\tilde b_R-\tilde s_R$- and $\tilde
\tau_L-\tilde\mu_L$-mixing which can then induce $b\to s$ and $\tau\to\mu$ transitions via SUSY loops. For a more detailed
derivation starting from an SO(10) superpotential see \cite{Girrbach:2011an}.
In a weak basis with diagonal up-type Yukawa matrix we
have
\begin{align}
\mathsf Y_d = \mathsf Y_\ell^\top =V_\text{CKM}^\star 
\begin{pmatrix}
 y_d & 0 & 0 \\
0 & y_s & 0 \\
0 &0 & y_b
\end{pmatrix}
U_D\,,\qquad   U_D = U_\text{PMNS}^\ast\,\text{diag}(1,\, e^{i\xi},\,1)
\end{align}
and the right-handed down squark mass matrix at the low scale reads
\begin{align}
 m_{\tilde{d}}^2(M_Z) =
\textrm{diag}\left(m_{\tilde{d}_1}^2,m_{\tilde{d}_1}^2,m_{\tilde{d}_1}^2
\left(1-\Delta_{\tilde{d}}\right)\right)\,,
\end{align}
where $\Delta_{\tilde{d}}\in [0,\,1]$ defines the relative mass splitting between the 1$^\text{st}$/2$^\text{nd}$ and
3$^\text{rd}$ down-squark
generation. It is generated by RG effects of the top Yukawa coupling and can reach $0.4$. Thus the CMM model shares the feature
of $U(2)^3$ models of heavy 1$^\text{st}$/2$^\text{nd}$ squark generations but a light 3$^\text{rd}$ generation. If we rotate to
mass eigenstate
basis and diagonalize $\mathsf Y_d$ the neutrino mixing enters $m_{\tilde{D}}^2$:
\begin{align}\label{equ:msquark}
 m_{\tilde{D}}^2 = U_D m_{\tilde{d}}^2 U_D^\dagger\approx
m_{\tilde{d}_1}^2\left(\begin{array}{ccc}
1 & 0 & 0\\
0 & 1-\frac{1}{2}\Delta_{\tilde{d}} &
-\frac{1}{2}\Delta_{\tilde{d}}e^{i\xi}\\
0 & -\frac{1}{2}\Delta_{\tilde{d}}e^{-i\xi}&
1-\frac{1}{2}\Delta_{\tilde{d}}
                                \end{array}
\right)\,.
\end{align}
Consequently, the 23-entry $ \propto\Delta_{\tilde{d}}$ is responsible for $\tilde b_R-\tilde s_R$-mixing and a
new
CP violating phase $\xi$ enters that affects $B_s-\overline{B}_s$-mixing. The ``$\approx$'' sign in (\ref{equ:msquark}) gets a
``$=$'' if one uses
tribimaximal mixing in $U_\text{PMNS}$. However, the latest data show that the reactor neutrino mixing angle $\theta_{13}$
is indeed non-zero \cite{Abe:2011fz,An:2012eh,Ahn:2012nd}. Including $\theta_{13}\neq$
the 12- and 13-entry in (\ref{equ:msquark}) are no longer zero, but still much smaller than the 23-entry. This gives small
corrections to $K-\overline{K}$- and $B_d-\overline{B}_d$-mixing.

\subsection{Phenomenology}

Four our global flavour analysis only seven parameters of the CMM model are relevant: the universal scalar soft mass $m_0$ and
trilinear coupling
$a_0$ at the Planck scale, the gluino mass $m_{\tilde g}$, the $D$-term mass splitting $D$, the phase of $\mu$, the phase $\xi$
and $\tan\beta$ (but the range $2.7\lesssim\tan\beta\lesssim 10$ follows from the superpotential and the requirement of
perturbative couplings up to the Planck scale). Similar to the constrained MSSM, the CMM model shares the nice feature of having
only a few model parameters, however the flavour structure is different: In the CMM model flavour universality is present at
$M_\text{Pl}$ but already broken at $M_\text{GUT}$ and hadronic and leptonic observables are correlated due to GUT boundary
conditions.

Flavour processes where we expect large CMM contributions are $B_s-\overline{B}_s$ mixing, $b\to s\gamma$ and $\tau\to\mu\gamma$
since here the neutrino mixing angle $\theta_{23}\approx 45^\circ$ connects the 2$^\text{nd}$ and 3$^\text{rd}$ generation.
CMM effects in $\mathcal{B}(B_s\to\mu^+\mu^-)$ are however small because at the electroweak scale the CMM model is a special
version of the MSSM  with small $\tan\beta$.
Consequently the CMM model is still compatible with new LHCb bound.
Due to the structure of (\ref{equ:msquark}) the contributions to $K-\overline{K}$ mixing, $B_d-\overline{B}_d$ mixing and $\mu\to
e\gamma$ are absent. However there are two sources of small corrections: a non-vanishing $\theta_{13}$ as
already mentioned and corrections due to dimension-5-Yukawa terms that are needed to fix  $\mathsf Y_d = \mathsf Y_\ell^\top$  for
the 1$^\text{st}$ and 2$^\text{nd}$ second generation. The latter point was worked out in \cite{Trine:2009ns}. In
\cite{Trine:2009ns} it was also shown that the tension in the SM between $\sin2\beta$ predicted from $|\varepsilon_K|$ and
$\Delta M_s/\Delta M_d$, and its direct measurement from $S_{\psi K_S}$ can be removed with the help of higher-dimensional Yukawa
couplings.

Results from our global analysis are the following: $\tau\to \mu\gamma$ constrains the sfermion masses of the first two
generations to lie above 1~TeV while the third generation can be much lighter. The sfermion masses can also be constrained by
$b\to s \gamma$ but  $\tau\to\mu\gamma$ gives stronger bounds. Gauginos can still be lighter. The lightest supersymmetric
particle is in most of the CMM parameter space the lightest neutralino with masses of $\mathcal{O}(100)~$GeV. Concerning
$B_s-\overline{B}_s$ mixing the situation shifted after the LHCb data for $S_{\psi\phi}$. Due to the free phase $\xi$ it is
possible to get large CP violation in the $B_s$ system in the CMM model while at the same time $\Delta M_s$ stays within its
experimental range.   In view of the data from CDF and D\O\ on $S_{\psi\phi}$ this property was very welcomed in 2010.
The new data on $S_{\psi\phi}$ implies new constraints on the model parameters, especially  on $\xi$ and on the ratio of gluino
and squark masses $m_{\tilde g}/M_{\tilde q}$ which must now be smaller than before. This was exemplarily shown in
\cite{Buras:2012ts}.  Consequently one previous advantage of the CMM model  over the constrained MSSM~-- the ability to generate a
large $S_{\psi\phi}$~-- is now gone. 

\begin{figure}[!tb]
\centering
 \includegraphics[width = 0.8\textwidth]{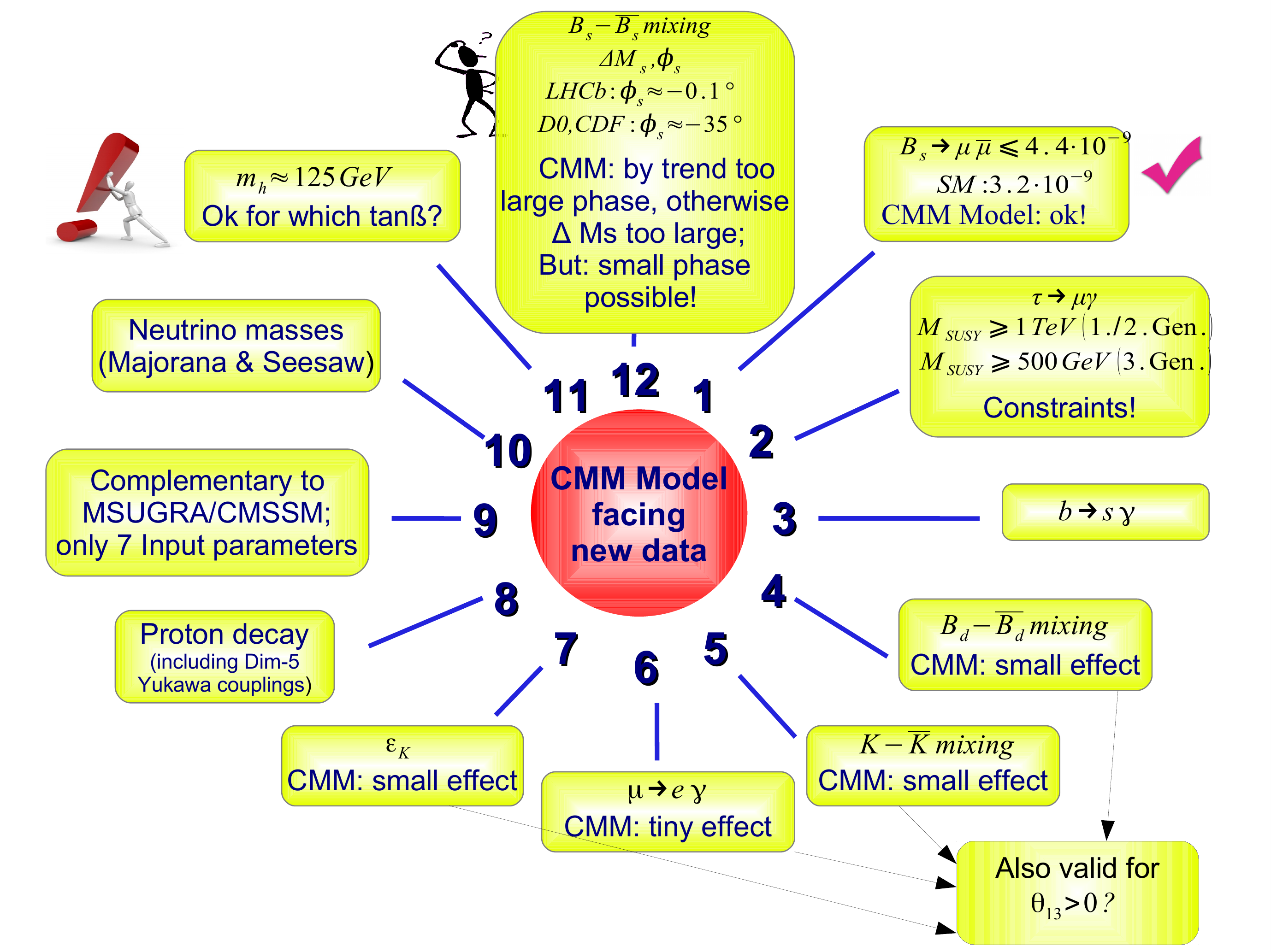}
\caption{Basic properties of the CMM model.}\label{fig:CMMUhr}~\\\hrule
\end{figure}

Another observable that needs further investigation is the Higgs mass. In the CMM model the mass of the lightest neutral Higgs  is
very sensitive to $\tan\beta$. Decreasing $\tan\beta$ also decreases the Higgs mass because  a larger top yukawa coupling
increases the mass splitting $\Delta_{\tilde d}$ in the renormalization group running  which leads to smaller stop
masses. Therefore the correction to the tree level Higgs mass in the MSSM gets smaller. In \cite{Girrbach:2011an} we pointed out
that $\tan\beta = 3$ is already excluded because in the regions where all flavour constraints are fulfilled the lightest Higgs
mass exceeds the LEP bound. For  $\tan\beta = 6$ the Higgs mass can be up to 120~GeV in the parameter range consistent with
flavour observables. Consequently one has to increase $\tan\beta$ further to accommodate a Higgs mass of 125~GeV.

\section{Summary}

In the first part we studied and compared correlations of $\Delta F = 2$ observables in CMFV and in a minimal version of
models with an approximate global  $U(2)^3$ flavour symmetry. These $MU(2)^3$ models are very simple non-MFV extensions
of the SM that avoid the  $\Delta M_{s,d}-\varepsilon_K$ tension present in CMFV.  We pointed out a triple correlation between
$S_{\psi\phi}$, $S_{\psi K_S}$ and $|V_{ub}|$ that constitutes an important test for $MU(2)^3$ models. A negative $S_{\psi\phi}$
could still be accommodated if the exclusive value of $|V_{ub}|$ turns out to be true. However than an 25\% enhancement in
$|\varepsilon_K|$ is needed. In the last part a concrete SO(10) SUSY GUT model, the CMM model was under consideration.
Instead of a written summary of the CMM model I refer to  Fig.~\ref{fig:CMMUhr} where the most important facts are listed.

\section*{Acknowledgments}

I thank the organizers of FLASY12 for the opportunity to give this talk and my collaborators  A.~Buras,  S.~J\"ager, M.~Knopf,
W.~Martens, U.~Nierste, C.~Scherrer and
S.~Wiesenfeldt for an enjoyable collaboration. I thank A.~Buras for  proofreading
this manuscript. I acknowledge financial support by  ERC Advanced Grant project ``FLAVOUR''(267104).


\bibliography{girrbach}

\providecommand{\href}[2]{#2}\begingroup\raggedright\begin{thebibliography}{10}

\bibitem{Aaij:2012ac}
{\bf LHCb collaboration} Collaboration, R.~Aaij {\em et.~al.}, {\it {Strong
  constraints on the rare decays $B_s \to \mu^+ \mu^-$ and $B^0 \to \mu^+
  \mu^-$}},  \href{http://xxx.lanl.gov/abs/1203.4493}{{\tt arXiv:1203.4493}}.

\bibitem{Buras:2012ts}
A.~J. Buras and J.~Girrbach, {\it {BSM models facing the recent LHCb data: A
  First look}},  {\em Acta Phys.Polon.} {\bf B43} (2012) 1427,
  [\href{http://xxx.lanl.gov/abs/1204.5064}{{\tt arXiv:1204.5064}}].

\bibitem{LHCb:2012py}
{\bf LHCb collaboration} Collaboration, B.~Bharucha {\em et.~al.}, {\it
  {Implications of LHCb measurements and future prospects}},
  \href{http://xxx.lanl.gov/abs/1208.3355}{{\tt arXiv:1208.3355}}.

\bibitem{Buras:2012ru}
A.~J. Buras, J.~Girrbach, D.~Guadagnoli, and G.~Isidori, {\it {On the Standard
  Model prediction for $\mathcal{B}(B_{s,d} \to \mu^+\mu^-)$}},
  \href{http://xxx.lanl.gov/abs/1208.0934}{{\tt arXiv:1208.0934}}.

\bibitem{deBruyn:2012wj}
K.~de~Bruyn, R.~Fleischer, R.~Knegjens, P.~Koppenburg, M.~Merk, {\em et.~al.},
  {\it {On Branching Ratio Measurements of $B_s$ Decays}},
  \href{http://xxx.lanl.gov/abs/1204.1735}{{\tt arXiv:1204.1735}}.

\bibitem{deBruyn:2012wk}
K.~de~Bruyn, R.~Fleischer, R.~Knegjens, P.~Koppenburg, M.~Merk, {\em et.~al.},
  {\it {A New Window for New Physics in $B^0_s\to \mu^+\mu^-$}},
  \href{http://xxx.lanl.gov/abs/1204.1737}{{\tt arXiv:1204.1737}}.

\bibitem{Clarke:1429149}
P.~Clarke, {\it Results on cp violation in $b_s$ mixing},
  \href{http://xxx.lanl.gov/abs/http://cdsweb.cern.ch/record/1429149/files/LHC%
b-TALK-2012-029.pdf}{{\tt
  http://cdsweb.cern.ch/record/1429149/files/LHCb-TALK-2012-029.pdf}}.

\bibitem{Aaij:2012qe}
{\bf LHCb Collaboration} Collaboration, R.~Aaij {\em et.~al.}, {\it {First
  evidence of direct CP violation in charmless two-body decays of Bs mesons}},
  {\em Phys.Rev.Lett.} {\bf 108} (2012) 201601,
  [\href{http://xxx.lanl.gov/abs/1202.6251}{{\tt arXiv:1202.6251}}].

\bibitem{LHCb-CONF-2012-002}
{\it Tagged time-dependent angular analysis of $b_s^0 \to j/\psi\phi$ decays at
  lhcb}, . Linked to LHCb-ANA-2012-004.

\bibitem{Lunghi:2008aa}
E.~Lunghi and A.~Soni, {\it {Possible Indications of New Physics in
  $B_d$-mixing and in $\sin(2 \beta)$ Determinations}},  {\em Phys. Lett.} {\bf
  B666} (2008) 162--165, [\href{http://xxx.lanl.gov/abs/0803.4340}{{\tt
  arXiv:0803.4340}}].

\bibitem{Buras:2008nn}
A.~J. Buras and D.~Guadagnoli, {\it {Correlations among new CP violating
  effects in $\Delta F = 2$ observables}},  {\em Phys. Rev.} {\bf D78} (2008)
  033005, [\href{http://xxx.lanl.gov/abs/0805.3887}{{\tt arXiv:0805.3887}}].

\bibitem{Nakamura:2010zzi}
{\bf Particle Data Group} Collaboration, K.~Nakamura {\em et.~al.}, {\it
  {Review of particle physics}},  {\em J.Phys.G} {\bf G37} (2010) 075021.

\bibitem{Brod:2011ty}
J.~Brod and M.~Gorbahn, {\it {Next-to-Next-to-Leading-Order Charm-Quark
  Contribution to the CP Violation Parameter $\epsilon_K$ and $\Delta M_K$}},
  {\em Phys.Rev.Lett.} {\bf 108} (2012) 121801,
  [\href{http://xxx.lanl.gov/abs/1108.2036}{{\tt arXiv:1108.2036}}].

\bibitem{Altmannshofer:2009ne}
W.~Altmannshofer, A.~J. Buras, S.~Gori, P.~Paradisi, and D.~M. Straub, {\it
  {Anatomy and Phenomenology of FCNC and CPV Effects in SUSY Theories}},  {\em
  Nucl.Phys.} {\bf B830} (2010) 17--94,
  [\href{http://xxx.lanl.gov/abs/0909.1333}{{\tt arXiv:0909.1333}}].

\bibitem{Aubert:2009wt}
{\bf BABAR Collaboration} Collaboration, B.~Aubert {\em et.~al.}, {\it {A
  Search for $B^+ \to \ell^+ \nu_{\ell}$ Recoiling Against $B^{-}\to D^{0}
  \ell^{-}\bar{\nu} X$}},  {\em Phys.Rev.} {\bf D81} (2010) 051101,
  [\href{http://xxx.lanl.gov/abs/0912.2453}{{\tt arXiv:0912.2453}}].

\bibitem{Ikado:2006un}
{\bf Belle Collaboration} Collaboration, K.~Ikado {\em et.~al.}, {\it {Evidence
  of the Purely Leptonic Decay $B^- \to\tau^-\bar\nu_\tau$}},  {\em
  Phys.Rev.Lett.} {\bf 97} (2006) 251802,
  [\href{http://xxx.lanl.gov/abs/hep-ex/0604018}{{\tt hep-ex/0604018}}].

\bibitem{HFAG}
{\bf Heavy Flavour Averaging Group} Collaboration, HFAG, {\it {Update of
  radiative and leptonic decays}}, .
  http://www.slac.stanford.edu/xorg/hfag/rare/2011/radll.

\bibitem{Nierste:2011na}
U.~Nierste, {\it {Flavour physics, supersymmetry and grand unification}},
  \href{http://xxx.lanl.gov/abs/1107.0621}{{\tt arXiv:1107.0621}}.

\bibitem{Lees:2012ju}
{\bf BABAR Collaboration} Collaboration, J.~Lees {\em et.~al.}, {\it {Evidence
  of $B\to \tau \nu$ decays with hadronic $B$ tags}},
  \href{http://xxx.lanl.gov/abs/1207.0698}{{\tt arXiv:1207.0698}}.

\bibitem{Adachi:2012mm}
{\bf Belle Collaboration} Collaboration, I.~Adachi {\em et.~al.}, {\it
  {Measurement of $B^- \to \tau^- \bar{\nu}_\tau$ with a Hadronic Tagging
  Method Using the Full Data Sample of Belle}},
  \href{http://xxx.lanl.gov/abs/1208.4678}{{\tt arXiv:1208.4678}}.

\bibitem{Blanke:2006yh}
M.~Blanke and A.~J. Buras, {\it {Lower bounds on $\Delta M_{s,d}$ from
  constrained minimal flavour violation}},  {\em JHEP} {\bf 0705} (2007) 061,
  [\href{http://xxx.lanl.gov/abs/hep-ph/0610037}{{\tt hep-ph/0610037}}].

\bibitem{Buras:2000xq}
A.~J. Buras and R.~Buras, {\it {A Lower bound on $\sin$ 2 beta from minimal
  flavor violation}},  {\em Phys.Lett.} {\bf B501} (2001) 223--230,
  [\href{http://xxx.lanl.gov/abs/hep-ph/0008273}{{\tt hep-ph/0008273}}].

\bibitem{Pomarol:1995xc}
A.~Pomarol and D.~Tommasini, {\it {Horizontal symmetries for the supersymmetric
  flavor problem}},  {\em Nucl.Phys.} {\bf B466} (1996) 3--24,
  [\href{http://xxx.lanl.gov/abs/hep-ph/9507462}{{\tt hep-ph/9507462}}].

\bibitem{Barbieri:1995uv}
R.~Barbieri, G.~Dvali, and L.~J. Hall, {\it {Predictions from a U(2) flavor
  symmetry in supersymmetric theories}},  {\em Phys.Lett.} {\bf B377} (1996)
  76--82, [\href{http://xxx.lanl.gov/abs/hep-ph/9512388}{{\tt
  hep-ph/9512388}}].

\bibitem{Barbieri:2011ci}
R.~Barbieri, G.~Isidori, J.~Jones-Perez, P.~Lodone, and D.~M. Straub, {\it
  {U(2) and Minimal Flavour Violation in Supersymmetry}},  {\em Eur.Phys.J.}
  {\bf C71} (2011) 1725, [\href{http://xxx.lanl.gov/abs/1105.2296}{{\tt
  arXiv:1105.2296}}].

\bibitem{Barbieri:2011fc}
R.~Barbieri, P.~Campli, G.~Isidori, F.~Sala, and D.~M. Straub, {\it {B-decay
  CP-asymmetries in SUSY with a $U(2)^3$ flavour symmetry}},  {\em Eur.Phys.J.}
  {\bf C71} (2011) 1812, [\href{http://xxx.lanl.gov/abs/1108.5125}{{\tt
  arXiv:1108.5125}}].

\bibitem{Barbieri:2012uh}
R.~Barbieri, D.~Buttazzo, F.~Sala, and D.~M. Straub, {\it {Flavour physics from
  an approximate $U(2)^3$ symmetry}},
  \href{http://xxx.lanl.gov/abs/1203.4218}{{\tt arXiv:1203.4218}}.

\bibitem{Barbieri:2012bh}
R.~Barbieri, D.~Buttazzo, F.~Sala, and D.~M. Straub, {\it {Less Minimal Flavour
  Violation}},  \href{http://xxx.lanl.gov/abs/1206.1327}{{\tt
  arXiv:1206.1327}}.

\bibitem{Crivellin:2011fb}
A.~Crivellin, L.~Hofer, and U.~Nierste, {\it {The MSSM with a softly broken
  $U(2)^3$ flavor symmetry}},  \href{http://xxx.lanl.gov/abs/1111.0246}{{\tt
  arXiv:1111.0246}}.

\bibitem{Crivellin:2011sj}
A.~Crivellin, L.~Hofer, U.~Nierste, and D.~Scherer, {\it {Phenomenological
  consequences of radiative flavor violation in the MSSM}},  {\em Phys.Rev.}
  {\bf D84} (2011) 035030, [\href{http://xxx.lanl.gov/abs/1105.2818}{{\tt
  arXiv:1105.2818}}].

\bibitem{Crivellin:2008mq}
A.~Crivellin and U.~Nierste, {\it {Supersymmetric renormalisation of the CKM
  matrix and new constraints on the squark mass matrices}},  {\em Phys.Rev.}
  {\bf D79} (2009) 035018, [\href{http://xxx.lanl.gov/abs/0810.1613}{{\tt
  arXiv:0810.1613}}].

\bibitem{Salatalk}
F.~Sala, {\it Talk given at flasy12: Flavour physics from an approximate
  $u(2)^3$ symmetry}, . Dortmund.

\bibitem{Buras:2012sd}
A.~J. Buras and J.~Girrbach, {\it {On the Correlations between Flavour
  Observables in Minimal U(2)$^3$ Models}},
  \href{http://xxx.lanl.gov/abs/1206.3878}{{\tt arXiv:1206.3878}}.

\bibitem{Chang:2002mq}
D.~Chang, A.~Masiero, and H.~Murayama, {\it {Neutrino mixing and large CP
  violation in $B$ physics}},  {\em Phys.Rev.} {\bf D67} (2003) 075013,
  [\href{http://xxx.lanl.gov/abs/hep-ph/0205111}{{\tt hep-ph/0205111}}].

\bibitem{Moroi:2000tk}
T.~Moroi, {\it {CP violation in $B_d \to \phi K_S$ in SUSY GUT with
  right-handed neutrinos}},  {\em Phys.Lett.} {\bf B493} (2000) 366--374,
  [\href{http://xxx.lanl.gov/abs/hep-ph/0007328}{{\tt hep-ph/0007328}}].

\bibitem{Girrbach:2011an}
J.~Girrbach, S.~Jager, M.~Knopf, W.~Martens, U.~Nierste, {\em et.~al.}, {\it
  {Flavor Physics in an SO(10) Grand Unified Model}},  {\em JHEP} {\bf 1106}
  (2011) 044, [\href{http://xxx.lanl.gov/abs/1101.6047}{{\tt
  arXiv:1101.6047}}].

\bibitem{Girrbach:2011wt}
J.~Girrbach, {\it {Flavour Physics in an SO(10) Grand Unified Model}},  {\em
  PoS} {\bf EPS-HEP2011} (2011) 183,
  [\href{http://xxx.lanl.gov/abs/1108.4852}{{\tt arXiv:1108.4852}}].

\bibitem{Abe:2011fz}
{\bf DOUBLE-CHOOZ Collaboration} Collaboration, Y.~Abe {\em et.~al.}, {\it
  {Indication for the disappearance of reactor electron antineutrinos in the
  Double Chooz experiment}},  {\em Phys.Rev.Lett.} {\bf 108} (2012) 131801,
  [\href{http://xxx.lanl.gov/abs/1112.6353}{{\tt arXiv:1112.6353}}].

\bibitem{An:2012eh}
{\bf DAYA-BAY Collaboration} Collaboration, F.~An {\em et.~al.}, {\it
  {Observation of electron-antineutrino disappearance at Daya Bay}},  {\em
  Phys.Rev.Lett.} {\bf 108} (2012) 171803,
  [\href{http://xxx.lanl.gov/abs/1203.1669}{{\tt arXiv:1203.1669}}].

\bibitem{Ahn:2012nd}
{\bf RENO collaboration} Collaboration, J.~Ahn {\em et.~al.}, {\it {Observation
  of Reactor Electron Antineutrino Disappearance in the RENO Experiment}},
  {\em Phys.Rev.Lett.} {\bf 108} (2012) 191802,
  [\href{http://xxx.lanl.gov/abs/1204.0626}{{\tt arXiv:1204.0626}}].

\bibitem{Trine:2009ns}
S.~Trine, S.~Westhoff, and S.~Wiesenfeldt, {\it {Probing Yukawa Unification
  with $K$ and $B$ Mixing}},  {\em JHEP} {\bf 0908} (2009) 002,
  [\href{http://xxx.lanl.gov/abs/0904.0378}{{\tt arXiv:0904.0378}}].

\end{thebibliography}\endgroup
\bibliographystyle{JHEP}

\end{document}